\documentclass[10pt,journal,a4paper,twoside]{./IMEJ}
\usepackage[pdftex]{graphicx}

\usepackage{textcomp}

\usepackage{amsmath}
\usepackage{booktabs}
\usepackage{threeparttable}
\usepackage[caption=false,font=footnotesize]{subfig}

\usepackage{hyperref}
\usepackage{url}
\hypersetup{
    colorlinks=true,
    citecolor=blue,
    linkcolor=blue,
    urlcolor=blue,
    }
\usepackage{balance}

\usepackage[capitalize]{cleveref}
\usepackage{xcolor}
\usepackage{enumitem}
\usepackage{booktabs}
\usepackage{multirow}

\usepackage[bibstyle=ieee, date=year, citestyle=numeric-comp, url=false, natbib=true, mincitenames=1, maxcitenames=2, uniquelist=false, uniquename = true]{biblatex}
\AtEveryBibitem{%
    \clearfield{doi}%
    \clearfield{note}%
    \clearlist{language}%
    \clearfield{issn}%
    \clearfield{isbn}%
    \clearfield{publisher}%
    \clearfield{editor}%
}
\begin{document}
%
%
\title{Experimental techniques for evaluating the performance of high-blockage cross-flow turbine arrays}
%
%
%

\author{Aidan Hunt
        and~Brian Polagye
\thanks{Pre-print submitted to the 15\textsuperscript{th} European Wave and Tidal Energy Conference. The version that appeared in the conference proceedings can be found at \url{https://doi.org/10.36688/ewtec-2023-203}.}%
\thanks{This work was supported by the United States Advanced Research Projects Agency – Energy (ARPA-E) under award number DE-AR0001441.}
\thanks{A. Hunt and B. Polagye are with the Department of Mechanical Engineering at the University of Washington, 3900 E Stevens Way NE, Seattle, WA 98195, U.S.A (e-mail: ahunt94@uw.edu).}
} 

%
%


\maketitle

\begin{abstract}
    In confined flows, such as river or tidal channels, arrays of turbines can convert both the kinetic and potential energy of the flow into renewable power. The power conversion and loading characteristics of an array in a confined flow is a function of the blockage ratio, defined as the ratio of the array's projected area to the channel cross-sectional area. In this work, we explore experimental methods for studying the effects of the blockage ratio on turbine performance while holding other variables constant. Two distinct methods are considered: one in which the array area is held constant and the channel area is varied, and another in which the array area is varied and the channel area is held constant. Using both approaches, the performance of a laboratory cross-flow turbine array in a water tunnel is evaluated at blockage ratios ranging from 30\% to 60\%. As the blockage ratio is increased, the coefficient of performance increases, eventually exceeding the Betz limit and unity. While similar trends are observed with both experimental approaches, at high blockage and high tip-speed ratios, the values of the performance and force coefficients are found to depend on the experimental approach. The advantages and disadvantages of each approach are discussed. Ultimately, we recommend investigating blockage effects using a fixed array area and variable channel area, as this approach does not convolve blockage effects with interactions between the turbine blades and support structures.
\end{abstract}

\begin{IMEJkeywords}
Cross-flow turbine, blockage, array, experiment
\end{IMEJkeywords}

\section{Introduction}
\IMEJPARstart{T}{he} efficiency of a turbine operating in a confined flow, such as a river or tidal channel, is influenced by how much of the channel the turbine occupies. The size of the turbine relative to the size of the channel is typically represented by the blockage ratio, defined as the ratio between the turbine's projected area and the channel cross-sectional area: 
\begin{equation}
    \beta = \frac{A_{turbine}}{A_{\mathrm{channel}}} .
\end{equation}
As the blockage ratio increases, the turbine presents greater resistance to the oncoming flow, and thus experiences greater thrust. For a constant volumetric flow rate, this thrust, combined with confinement from the channel boundaries, yields accelerated flow through the turbine rotor.
As a consequence, a turbine operating in a confined flow produces more power relative to the same turbine in an unconfined flow \citep{garrett_efficiency_2007, consul_blockage_2013, houlsby_power_2017}.

Lateral arrays or ``fences'' of turbines deployed in river or tidal channels can harness these blockage effects to enhance power production \citep{vennell_tuning_2010, nishino_efficiency_2012, houlsby_power_2017}.
However, since blockage-driven increases in power generation are accompanied by increases in the forces on the turbine, understanding how both power and loads scale with the blockage ratio is critical for rotor design \citep{schluntz_effect_2015} and control \citep{burton_wind_2011}
Additionally, the blockage ratio in natural channels may vary daily (e.g., with the tides), seasonally (e.g., runoff from snowmelt or storms), or as needed when individual turbines are deactivated for maintenance or to allow the passage of vessels and marine animals. Therefore, understanding how changes in blockage will alter turbine hydrodynamics, and thus power production, is necessary for the management of these systems.

The study of blockage effects is also applicable to turbines that are intended for use in unconfined environments. Many laboratory settings in which model turbines are tested, such as wind tunnels or flumes, are inherently confined flows, and the associated blockage effects will yield augmented performance \citep{bahaj_power_2007, ross_effects_2022}, even at blockages of 10\% and below \citep{battisti_aerodynamic_2011, dossena_experimental_2015}. Further, in experimental design, increasing model scale to achieve Reynolds numbers that are more representative of a full scale turbine generally increases blockage.
While analytical corrections have been developed to predict unconfined turbine performance using measurements of confined turbine performance \citep{glauert_airplane_1935, maskell_ec_theory_1963, houlsby_application_2008, whelan_free-surface_2009}, most are simplified models based on linear momentum theory, and their accuracy can vary \citep{ross_experimental_2020}. Given this, dedicated study of blockage effects, particularly at the upper end of achievable blockage in practical situations, is relevant.

\begin{table*}[]
    \caption{Summary of experimental studies that consider the effects of varying the blockage ratio for axial-flow turbines (AFTs) and cross-flow turbines (CFTs).}
    \label{tab:priorWork}
    \begin{threeparttable}
        \centering
        \resizebox{\textwidth}{!}{      
            \begin{tabular}{@{}lccccc@{}}
                \toprule
                Author & Year & Turbine Type & Facility Type & $\beta$ tested [\%] & $\beta$ varied by changing \\ \midrule
                \citet{whelan_free-surface_2009} & 2009 & AFT & Wind tunnel, water flume & 5, 64 & Test facility \\
                \citet{mcadam_experimental_2010} & 2010 & CFT & Water flume & 50, 62.5 & Water depth \\
                \citet{battisti_aerodynamic_2011} & 2011 & CFT & Wind tunnel with removable walls & 2.8\tnote{\dag}, 10 & Wind tunnel size \\
                \citet{chen_blockage_2011} & 2011 & AFT & Wind tunnel & 10.2, 20.2, 28.3 & Turbine diameter \\
                \citet{ross_wind_2011} & 2011 & CFT & Wind tunnel & 2, 3.5, 8 & Turbine scale \\
                \citet{burdett_scaling_2012} & 2012 & AFT & Wind tunnel & 19, 33.8, 52.8 & Turbine diameter \\
                \citet{birjandi_power_2013} & 2013 & CFT & Water flume & 24.5 - 49.2\tnote{\dag} & Water depth \\
                \citet{mctavish_experimental_2014} & 2014 & AFT & Water flume & 6.3, 9.9, 14.3, 19.4, 25.4 & Turbine diameter \\
                \citet{gaurier_tidal_2015} & 2015 & AFT & Multiple flumes and tow tanks & 1.2, 3.3, 4.8 & Test facility \\
                \citet{ryi_blockage_2015} & 2015 & AFT & Multiple wind tunnels & 8.1, 18.0, 48.1 & Test facility \\
                \citet{dossena_experimental_2015} & 2015 & CFT & Wind tunnel with removable walls & 2.8\tnote{\dag}, 10 & Wind tunnel size \\
                \citet{jeong_blockage_2018} & 2018 & CFT & Multiple wind tunnels & 3.5, 13.4, 24.7 & Test facility \\
                \citet{ross_experimental_2020} & 2020 & CFT, AFT & Multiple flumes & AFT: 2, 35. CFT: 3, 36 & Test facility \\
                \citet{ross_experimental_2020-1} & 2020 & CFT & Multiple flumes & 14, 36 & Test facility \\
                \citet{ross_effects_2022} & 2022 & CFT & Water tunnel & 9.9, 12.4 & Blade span \\ \bottomrule
            \end{tabular}
        }
            \begin{tablenotes}
                 \item[\dag] Value estimated from provided turbine and channel dimensions.
            \end{tablenotes}
        \end{threeparttable}
\end{table*}

\Cref{tab:priorWork} summarizes prior experimental work that has explored how changing the blockage ratio affects the performance of both axial-flow turbines and cross-flow turbines. While we focus this review on experimental studies, we acknowledge that there is a complementary body of numerical work (e.g., \citep{nishino_effects_2012, consul_blockage_2013, gauvin-tremblay_two-way_2020, gauvin-tremblay_hydrokinetic_2022}). In alignment with theory, an increase in blockage is found to increase the thrust loading on the turbine \citep{dossena_experimental_2015, ross_experimental_2020, ross_experimental_2020-1, ross_effects_2022} and the maximum efficiency of the turbine \citep{whelan_free-surface_2009, mcadam_experimental_2010, ross_wind_2011, birjandi_power_2013, ryi_blockage_2015, dossena_experimental_2015, jeong_blockage_2018, ross_experimental_2020, ross_experimental_2020-1, ross_effects_2022}, as well as the tip-speed ratio at which this maximum occurs.
From a fluid dynamic standpoint, an increase in blockage is observed to accelerate the flow through and around the rotor, as well as narrow the turbine wake \citep{ battisti_aerodynamic_2011, mctavish_experimental_2014, ross_experimental_2020-1}.
Although these key trends are common across these studies, multiple approaches for varying blockage have been used across these experiments. Some studies vary $\beta$ by changing the size of the channel; for example, testing the same turbine in different facilities \citep{whelan_free-surface_2009, ryi_blockage_2015, jeong_blockage_2018, ross_experimental_2020, ross_experimental_2020-1}, or altering the water depth in a flume \citep{mcadam_experimental_2010, birjandi_power_2013}. Others vary $\beta$ by changing the dimensions of the turbine itself \citep{chen_blockage_2011, ross_wind_2011, burdett_scaling_2012, mctavish_experimental_2014, ross_effects_2022}. 

All of these experimental approaches yield similar trends, yet differences in approach limit a deeper understanding of blockage effects.
While differences in rotor geometry \citep{schluntz_effect_2015, ross_experimental_2020} and type of test facility \citep{gaurier_tidal_2015} can influence observed trends, even blockage effects measured in a single test facility with a single turbine design can be inadvertently convolved with the effects of other variables. For example, \citet{mcadam_experimental_2010} and \citet{birjandi_power_2013} both change the blockage ratio by changing water depth in a flume, but in doing so convolve the effects of blockage with those of the Froude number and proximity to the free surface. Overall, the variety of approaches employed by prior work motivate a thorough consideration of experimental methods for studying blockage effects on turbines and the robustness of these techniques.

In this work, we discuss two approaches by which the blockage ratio of a turbine array can be varied in a single test facility while holding constant or minimizing the effects of other dimensionless parameters. Using both approaches, we evaluate the performance of a cross-flow turbine array at blockage ratios between 30\% and 60\%, and compare the results. These blockages are of importance for associated research employing control co-design to understand the potential for high-blockage arrays of cross-flow turbines to significantly reduce cost of energy.

\section{Background}
\label{sec:background}

Consider a row of identical, straight-bladed, vertical-axis cross-flow turbines operating in a rectangular water channel. Neglecting the area of any support structures or fixturing, the array blockage ratio for such a system is given by

\begin{equation}
    \beta = \frac{A_{\mathrm{turbines}}}{A_{\mathrm{channel}}} = \frac{NHD}{hw} \ \ , 
    \label{eq:blockage}
\end{equation}

\noindent where $N$ is the number of turbines, $D$ is the rotor diameter, $H$ is the blade span, $h$ is the time-varying channel depth measured at the turbine's axis of rotation, and $w$ is the channel width (\cref{fig:arrayDimensions}). To characterize the array's performance as a function of $\beta$, there are two distinct approaches by which $\beta$ can be varied.

\subsection*{Approach 1: Fixed $A_{\mathrm{turbines}}$, variable $A_{\mathrm{channel}}$}

\begin{figure}
    \centering
    \includegraphics[width=\linewidth]{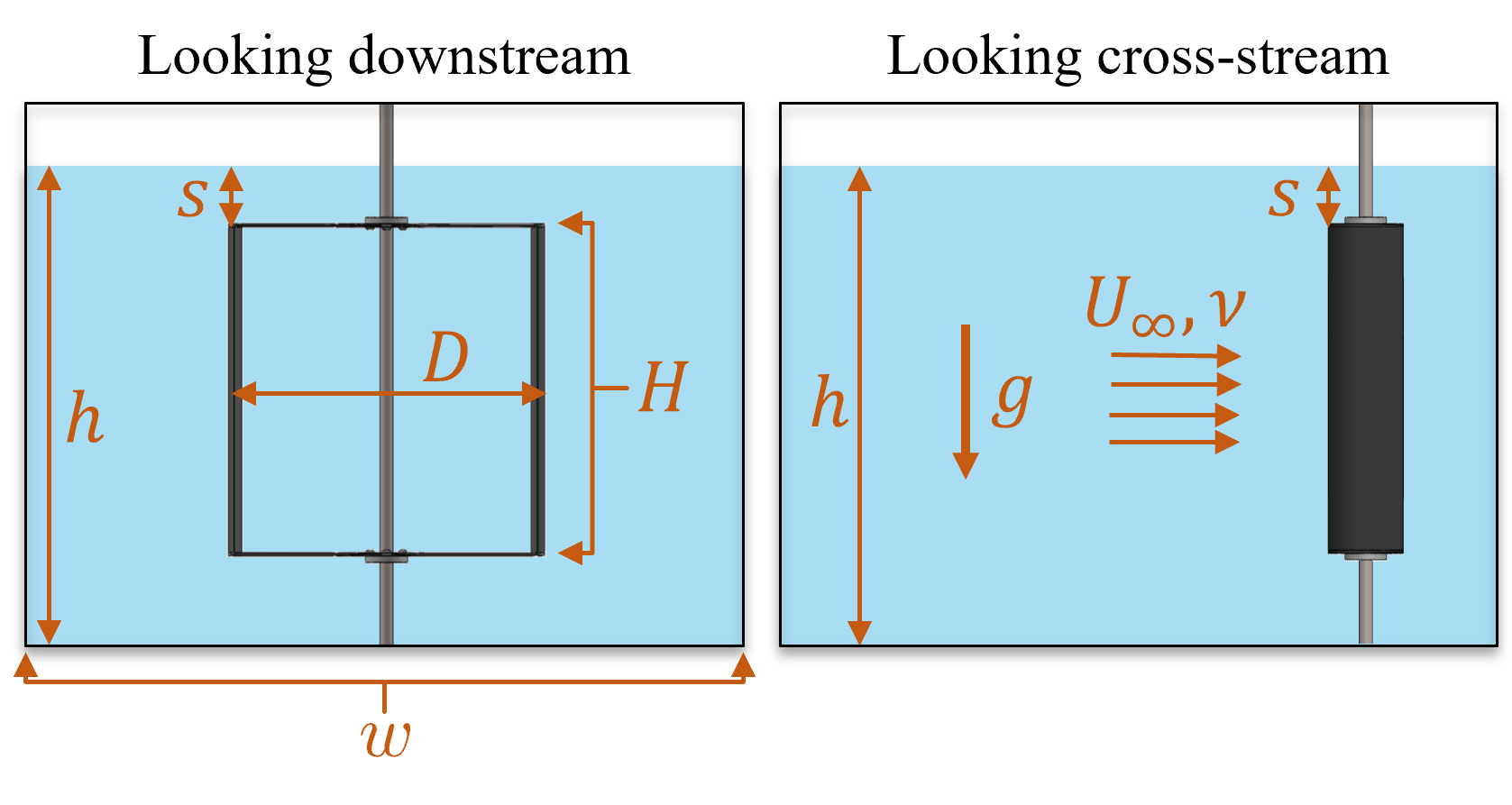}
    \caption{Dimensions and fluid properties for a single straight-bladed cross-flow turbine in a water channel, as viewed looking downstream (left) and looking cross-stream (right).}
    \label{fig:arrayDimensions}
\end{figure}

For a fixed array geometry, $\beta$ can be varied by changing the cross-sectional area of the channel. For both flumes and wind tunnels, it is possible, though logistically challenging, to reduce $A_{\mathrm{channel}}$ by changing channel width through the installation of intermediate partitions \citep{mcadam_experimental_2010, battisti_aerodynamic_2011, dossena_experimental_2015}. For flumes, $A_{\mathrm{channel}}$ is more easily altered by varying the water depth ($h$). However, to experimentally isolate the effects of blockage on array performance while doing so, other non-dimensional flow parameters must be carefully controlled. 

For example, consider an increase in $\beta$ achieved via a decrease in the water depth. This decrease in $h$ will simultaneously increase the depth-based Froude number,

\begin{equation}
    Fr_h = \frac{U_{\infty}}{\sqrt{gh}} \ \ , 
    \label{eq:depthFroude}
\end{equation}

\noindent where $U_{\infty}$ is the freestream velocity and $g$ is the acceleration due to gravity. The depth-based Froude number represents the balance between inertial forces in the flow and gravitational forces in the flow. 
The array's proximity to the free surface will also change as $h$ is decreased, which can be represented by the normalized submergence, $s/h$, where $s$ is the distance between the free surface and the top of the turbine rotors (\cref{fig:arrayDimensions}). Both $Fr_h$ and $s/h$ have been shown to impact turbine performance \citep{ross_effects_2022, birjandi_power_2013, kolekar_performance_2015, kolekar_blockage_2019}. Therefore, if $h$ decreases, $U_{\infty}$ must be decreased to hold $Fr_h$ constant. Similarly, $s$ must be decreased (i.e., the rotors positioned dimensionally nearer to the free surface) to hold $s/h$ constant at this new water depth.

However, a decrease in $U_{\infty}$ will also decrease the Reynolds number, which here is defined with respect to the turbine diameter as

\begin{equation}
    Re_D =  \frac{ U_{\infty} D}{\nu} \ \ ,
    \label{eq:ReD}
\end{equation}

\noindent where $\nu$ is the kinematic viscosity. The Reynolds number represents the balance between inertial forces and viscous forces in the flow. The dependence of turbine performance on the Reynolds number is well documented \citep{miller_vertical-axis_2018, miller_solidity_2021, bachant_effects_2016, ross_effects_2022}. Although turbine performance becomes independent of the Reynolds number above a certain threshold, for cross-flow turbines this is difficult to achieve at laboratory scale without the use of compressed-air wind tunnels \citep{miller_vertical-axis_2018}, so in most facilities, the Reynolds number must be held constant to isolate blockage effects. To compensate for the decrease in $U_{\infty}$ necessitated by holding $Fr_h$ constant, the kinematic viscosity $\nu$ can be increased by changing water temperature ($T$). In this way, $\beta$ may be varied in water channels, while holding $Fr_h$, $s/h$, and $Re_D$ constant, without the complications of intermediate partitions to adjust width. 

\subsection*{Approach 2: Variable $A_{\mathrm{turbines}}$, fixed $A_{\mathrm{channel}}$}

Alternatively, $\beta$ can be varied by changing $A_{\mathrm{turbines}}$. For an array of straight-bladed cross-flow turbines, this can be achieved by changing $N$, $H$, and/or $D$. However, as for the previous approach, experimental isolation of blockage effects can be complicated by unintended effects introduced by changing the turbine geometry or the number of turbines.

For example, if $A_{\mathrm{turbines}}$ is increased by increasing the rotor diameter, several geometric and flow parameters are simultaneously varied. First, increasing $D$ decreases the chord-to-radius ratio, $c/R$, which influences the flow curvature effects (e.g., virtual camber and virtual incidence) experienced by the blades \citep{migliore_flow_1980}. Second, increasing $D$ also increases the diameter-based Reynolds number $Re_D$ \eqref{eq:ReD}, and necessitates corresponding increases to $U_{\infty}$ and/or $\nu$ to hold $Re_D$ constant across blockage cases. Finally, for a fixed-width channel, changing $D$ changes both the spacing between adjacent turbines as well as the proximity of the turbines at the ends of the array to the channel side walls. For proximity on the order of the blade chord length, this alters the hydrodynamic interactions between adjacent rotors \citep{scherl_geometric_2020} as well as the lateral boundary effects on array performance \citep{gauvin-tremblay_two-way_2020}. 

One could also increase $A_{\mathrm{turbines}}$ by increasing the number of turbines on a cross-sectional transect. However, as for a change in $D$, for a fixed-width channel the proximity between turbines and the channel side-walls decreases, producing the same changes to boundary effects as increasing diameter. Further, this introduces the potential for new turbine-turbine interactions, which are similar to turbine-wall interactions, but have an additional degree of freedom in the rotational phase of adjacent turbines \citep{scherl_geometric_2020, gauvin-tremblay_hydrokinetic_2022}. As a result, if the number of turbines in the array is changed, it can be difficult to separate the effects of blockage from those of intra-array interactions and boundary effects. 

Conversely, if $A_{\mathrm{turbines}}$ is increased by increasing only $H$ (and each rotors' position in the water column is adjusted to hold the submergence depth $s$ constant), then only the rotor aspect ratio, $H/D$, is simultaneously varied with $\beta$. 
Given that this method varies fewer secondary parameters than either $D$ or $N$, changing $A_{\mathrm{turbines}}$ via $H$ alone is a conceptually attractive means of varying array blockage for fixed $A_{\mathrm{channel}}$. 
However, this approach would only be effective at experimentally isolating blockage effects if the effects of changing aspect ratio on turbine performance are minor by comparison. 
Prior work by \citet{hunt_effect_2020} has shown that the efficiency of a single turbine with blade-end struts at $\beta=11\%$ is invariant for $H/D = 0.95\!-\!1.63$, although the range of invariance likely depends on the type of support structure used (e.g., endplates, midspan struts) \citep{strom_impact_2018, villeneuve_increasing_2021} and may be different for high-blockage arrays. As $H/D$ is further decreased via a decrease in $H$, it is hypothesized that hydrodynamic interactions between the blades and the support structures could become more prominent and alter performance. 

In summary, for fixed $A_{\mathrm{turbines}}$, $\beta$ is best varied by changing the water depth in the channel, and for fixed $A_{\mathrm{channel}}$, $\beta$ is best varied by changing the height of the rotors via the blade span. In this work, both methods are evaluated experimentally and compared.

\section{Experimental Methods}
\label{sec:methods}

\begin{table*}[t]
    \centering
    \caption{Experimental parameters for each blockage condition tested.}
    \resizebox{\textwidth}{!}{      
        \begin{tabular}{@{}ccccccccccc@{}}
            \toprule
            \multicolumn{3}{c}{Target Blockage Condition} & \multicolumn{1}{l}{} & \multicolumn{4}{c}{Nominal Flume   Parameters} & \multicolumn{3}{c}{Nominal   Non-Dimensional Flow Parameters} \\ \cmidrule(r){1-3} \cmidrule(l){5-11} 
            $\beta$   (\%) & $A_{\mathrm{turbines}} \ (\mathrm{m}^2)$ & $A_{\mathrm{channel}} \ (\mathrm{m}^2)$ & $H$ (m) & $h$ (m) & $U_{\infty}$ (m/s) & $s$ (m) & $T$ [$^{\circ}\mathrm{C}$] & $s/h$ & $Fr_h$ & $Re_D$ \\ \midrule
            30.0 & \multirow{7}{*}{0.135} & 0.451 & \multirow{7}{*}{0.215} & 0.593 & 0.528 & 0.317 & 21.0 & 0.53 & \multirow{7}{*}{0.219} & \multirow{7}{*}{$1.7\!\times\!10^5$} \\
            33.4 &  & 0.406 &  & 0.534 & 0.501 & 0.257 & 23.3 & 0.48 &  &  \\
            36.7 &  & 0.369 &  & 0.485 & 0.477 & 0.209 & 25.4 & 0.43 &  &  \\
            40.1 &  & 0.338 &  & 0.445 & 0.457 & 0.169 & 27.3 & 0.38 &  &  \\
            45.0 &  & 0.301 &  & 0.396 & 0.431 & 0.120 & 30.1 & 0.30 &  &  \\
            50.0 &  & 0.271 &  & 0.356 & 0.409 & 0.080 & 32.6 & 0.22 &  &  \\
            55.0 &  & 0.246 &  & 0.324 & 0.390 & 0.048 & 35.0 & 0.15 &  &  \\ \midrule
            45.0 & \multirow{4}{*}{0.203} & 0.451 & \multirow{4}{*}{0.322} & 0.593 & 0.528 & 0.210 & 21.0 & 0.35 & \multirow{4}{*}{0.219} & \multirow{4}{*}{$1.7\!\times\!10^5$} \\
            50.0 &  & 0.406 &  & 0.534 & 0.501 & 0.150 & 23.3 & 0.28 &  &  \\
            55.0 &  & 0.369 &  & 0.485 & 0.477 & 0.102 & 25.4 & 0.21 &  &  \\
            60.0 &  & 0.338 &  & 0.445 & 0.457 & 0.062 & 27.3 & 0.14 &  &  \\ \bottomrule
        \end{tabular}
    }
    \label{tab:expMatrix}
\end{table*}

\subsection{Test Setup}

\begin{figure}[t]
    \centering
    \includegraphics[width=\linewidth]{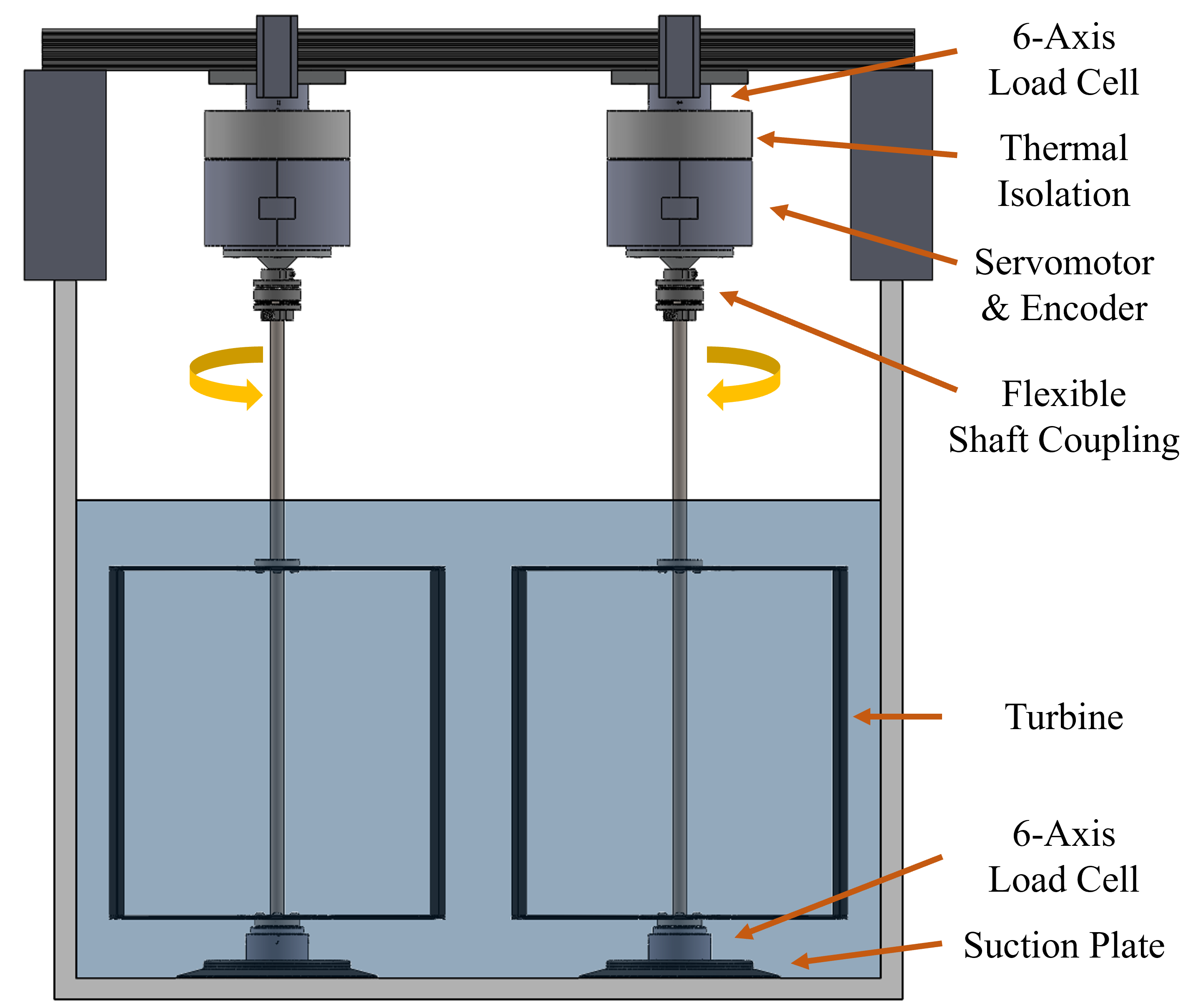}
    \caption{A rendering of the experimental test-rig, as viewed from upstream. The array shown is at $\beta=60\%$ with $H = 0.322 \ \mathrm{m}$, $h = 0.445 \ \mathrm{m}$, and $s = 0.062 \ \mathrm{m}$.}
    \label{fig:testRig}
\end{figure}

Experiments were conducted in the Alice C. Tyler recirculating water flume at the University of Washington. The flume has a test section that is 0.76 m wide and 4.88 m long, and can accommodate water depths up to 0.60 m and flow speeds up to ${\sim}\!1.1$ m/s. The flume is equipped with both a heater and chiller for temperature control, and can maintain water temperatures between $10^{\circ} \mathrm{C}$ and $40^{\circ}\mathrm{C}$ during operation.

The laboratory-scale array consists of two identical straight-bladed cross-flow turbines. The rotors are each two-bladed and have a diameter of 0.315 m, defined as the outermost circle swept by the turbine blades. Each blade has a NACA 0018 profile, a 0.0742 m chord length, and is mounted at a $6^{\circ}$ preset pitch angle as referenced from the quarter chord. The blades are attached to the central driveshaft of each rotor using thin, hydrodynamic blade-end struts (NACA 0008 profile, 0.0742 m chord length). The chord-to-radius ratio is 0.47 and the solidity is 0.15.

The two rotors are integrated into the experimental set-up shown in \cref{fig:testRig}, which consists of two identical test-rigs. The top of each turbine's central shaft is connected by a flexible shaft coupling (Zero-Max SC040R) to a servomotor (Yaskawa SGMCS-05BC341) which regulates the rotation rate of the turbine. The angular position of each turbine is measured via the servomotor encoder, from which the angular velocity is estimated. The bottom of each turbine's central shaft sits in a bearing. The net forces and torques on each turbine are measured by a pair of 6-axis load cells: an upper load cell (ATI Mini45-IP65) mounted to the servomotor and fixed to a crossbeam, and a lower load cell (ATI Mini45-IP68) mounted to the bottom bearing and fixed to the bottom of the flume via a suction plate. Measurements from the load cells and servomotor encoders for both turbines are acquired synchronously at 1000 Hz in MATLAB using a pair of National Instruments PCIe-6353 DAQs.

The freestream velocity is measured using an acoustic Doppler velocimeter (Nortek Vectrino Profiler) sampling at 16 Hz. The velocimeter sampled a single cell positioned laterally in the center of the flume, vertically at the array midplane, and 5 turbine diameters upstream of the array centerline. Velocity measurements are despiked using the method of \citet{goring_despiking_2002}. The water depth upstream of the array is measured at the center of the flume ${\sim}5.8$ turbine diameters upstream of the array centerline by an ultrasonic free-surface transducer (Omega LVU 32) sampling at 1 Hz. The water temperature is measured using a temperature probe (Omega Ultra-Precise RTD) and maintained within $\pm 0.1^{\circ} \mathrm{C}$ of the target value during each experiment.

\subsection{Test Matrix}

Array performance is characterized at blockage ratios ranging from $30\%$ to $60\%$ using combinations of $A_{\mathrm{turbines}}$ and $A_{\mathrm{channel}}$. \Cref{tab:expMatrix} summarizes the turbine geometries and flume conditions used to achieve each $\beta$. Using two different blade spans, two values of $A_{\mathrm{turbines}}$ are tested. For each value of $A_{\mathrm{turbines}}$, $\beta$ is varied by changing the water depth, with corresponding variations in $U_{\infty}$ and $\nu$ to maintain constant $Fr_h$ and $Re_D$ across all experiments. To test whether the same characteristic performance is measured for arrays with identical $\beta$, but different $A_{\mathrm{turbines}}$ and $A_{\mathrm{channel}}$, both the $A_{\mathrm{turbines}} = 0.135 \mathrm{\ m}^2$ rotors and the $A_{\mathrm{turbines}} = 0.203 \mathrm{\ m}^2$ rotors are tested at $45.0\%$, $50.0\%$, and $55.0\%$ blockage.

The values of $Fr_h$ and $Re_D$, which are held constant across all experiments, are constrained by the maximum $U_{\infty}$ at which the highest blockage arrays can be tested. This velocity, in turn, is constrained by rotor ventilation (i.e., air entrainment), the onset of which becomes more likely with decreasing $s/h$ and results in significant performance degradation due to an increase in form drag on the blades \citep{birjandi_power_2013, young_ventilation_2017}. As turbines at the highest blockages ($\beta \geq 55.0\%$) necessarily operate close to the free surface, the maximum allowable $U_{\infty}$ for these cases is set such that any ventilation occurs well beyond the optimal performance point. While $s/h$ could be held constant across all experiments by adjusting the array submergence depth, preliminary experiments showed that varying $s/h$ has minimal effect on array performance until ventilation begins to occur. Therefore, we instead choose to maximize $s/h$ at each blockage so as to limit the risk of ventilation as much as possible. 

\subsection{Array Layout and Control}
\label{sec:arrayLayout}

\begin{figure}
    \centering
    \includegraphics[width=\linewidth]{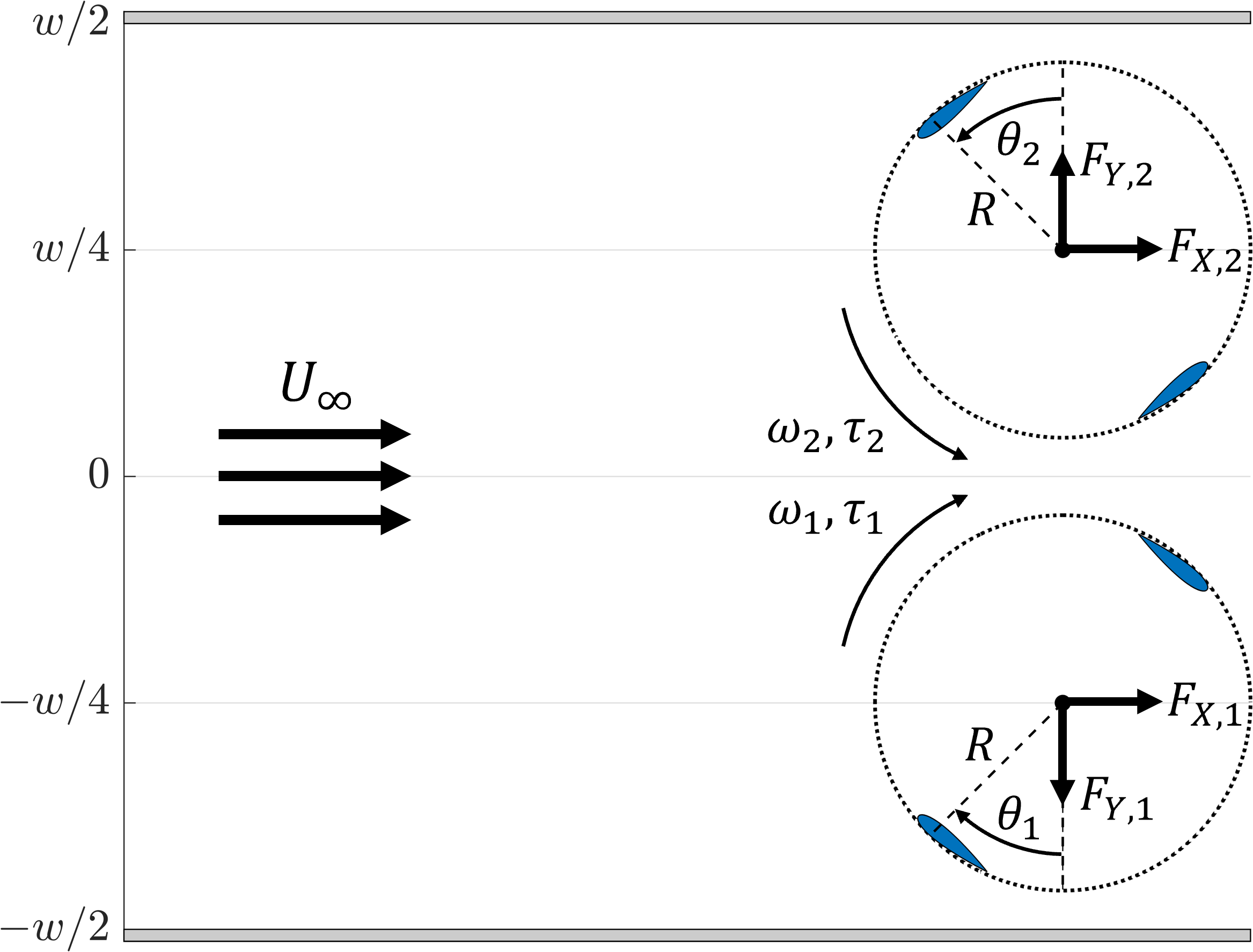}
    \caption{Overhead view of the array layout in the Tyler flume, with key measured quantities annotated.}
    \label{fig:arrayOverhead}
\end{figure}

An overhead view of the array layout is shown in \cref{fig:arrayOverhead}. The center-to-center spacing between the turbines is ${\sim}1.2D$, and the array is positioned laterally such that the blade-to-blade spacing between adjacent turbines is twice the wall-to-blade spacing (i.e., the walls notionally correspond to symmetry planes in a larger array). The turbines in the array were operated under a counter-rotating, phase-locked scheme, wherein both turbines rotate at the same, constant speed, but in opposite directions, with a constant angular phase offset, $\Delta \theta$, between them. This control strategy was achieved by specifying the angular velocities of the rotors, which yields similar time-average performance to an array with a constant control torque \citep{polagye_comparison_2019}. The turbines were counter-rotated such that the blades of adjacent rotors pass nearest each other while moving downstream, which has been shown to augment performance relative to other rotation schemes \citep{scherl_optimization_2022, gauvin-tremblay_hydrokinetic_2022}. We limit the present experiments to $\Delta \theta = 0^{\circ}$, an operating case in which the lateral forces and reaction torques for a pair of counter-rotating turbines are equal and opposite. A closed-loop controller maintained $\Delta \theta$ to within $1^{\circ}$ of the target value at all rotation rates across all experiments.

\subsection{Performance Metrics}
\label{sec:perfMet}

Performance metrics are calculated for individual turbines from the measured quantities shown in \cref{fig:arrayOverhead}. The rotation rate, which is the same for both turbines, is non-dimensionalized as the ratio of the blade tangential velocity to the freestream velocity, or the tip-speed ratio 

\begin{equation}
    \lambda = \frac{\omega R}{U_{\infty}} \ \ ,
    \label{eq:TSR}
\end{equation}

\noindent where $\omega$ is the angular velocity of the turbine and $R$ is the turbine radius. Data are collected at each tip-speed ratio for 60 seconds, and the time series is cropped to an integer number of turbine rotations before performance metrics are calculated.

The efficiency (formally, the coefficient of performance) of each turbine is the mechanical power produced normalized by the kinetic power in the freestream flow that passes through the turbine's projected area

\begin{equation}
    C_{P,i} = \frac{\tau_{i}\omega_{i}}{\frac{1}{2}\rho U_{\infty}^3 D H} \ \ ,
    \label{eq:cp}
\end{equation}

\noindent where $\omega_{i}$ and $\tau_{i}$ are the angular velocity and hydrodynamic torque on turbine $i$, and $\rho$ is the density of the working fluid.
The efficiency of each turbine is a function of the power produced by its blades and power losses due to parasitic torque on its blade support structures; in this case, blade-end struts. Because a constant set of blade-end struts is used for all experiments (i.e., the thickness of the struts is not scaled as the blade span is changed), the relative impact of these parasitic torques on turbine efficiency is larger for turbines with shorter blades (i.e., smaller $A_{\mathrm{turbines}}$ as in \Cref{tab:expMatrix}). To account for this, we utilize the approach of \citet{bachant_experimental_2016, strom_impact_2018} to estimate a blade-level $C_P$ for each turbine (i.e., the efficiency of the turbine blades in the absence of the support structures) via superposition as
\begin{equation}
    \begin{split}
        C_{P,i,\mathrm{blade}}(\beta, \lambda) & \approx C_{P,i,\mathrm{turbine}}(\beta, \lambda) \\ & - C_{P,i,\mathrm{supports}}(\beta, \lambda)
    \end{split}
    \label{eq:cpBlade}
\end{equation}

\noindent where $C_{P,i,\mathrm{turbine}}$ is the measured efficiency of the full turbine $i$, and $C_{P,i,\mathrm{supports}}$ is the measured efficiency of turbine $i$ with no blades attached.

Structural loads on each turbine are characterized via the thrust and lateral force coefficients, respectively given as

\begin{equation}
    C_{F_{X},i} = \frac{F_{X,i}}{\frac{1}{2} \rho U_{\infty}^2 DH} \ \ ,
    \label{eq:cThrust}
\end{equation}
\begin{equation}
    C_{F_{Y},i} = \frac{F_{Y,i}}{\frac{1}{2} \rho U_{\infty}^2 DH} \ \ ,
    \label{eq:cLat}
\end{equation}

\noindent where $F_{X,i}$ and $F_{Y,i}$ are the streamwise force and lateral force, respectively, on turbine $i$. To estimate blade-level loading for each turbine in the array, we apply superposition equations for $C_{F_{X},i}$ and $C_{F_{Y},i}$ analogous to that given for $C_{P,i}$ in \eqref{eq:cpBlade}. However, we note that, unlike for $C_{P,i}$, the validity of this approach for estimating blade-level $C_{F_{X},i}$ and $C_{F_{Y},i}$ has not been examined in the existing literature.

Since the turbines in this array are identical, the array-average performance metrics are obtained simply as the average of the individual turbine performance metrics. For example, $C_{P,array}$ is simply the average of $C_{P,1}$ and $C_{P,2}$ for the full turbine, and the average of $C_{P,1,\mathrm{blade}}$ and $C_{P,2,\mathrm{blade}}$ at the blade-level.
As noted in \Cref{sec:arrayLayout}, the net lateral force on this array of two identical counter-rotating turbines is zero since when $\Delta \theta = 0^{\circ}$, the array is symmetric about its centerline. However, by defining the directions of $F_{Y,1}$ and $F_{Y,2}$ with this counter-rotation in mind as in \cref{fig:arrayOverhead}, the array-average of $C_{F_{Y}}$ is nonzero and represents the average lateral force coefficient experienced by an individual rotor.

\section{Results and Discussion}
\label{sec:results}

\subsection{Non-dimensional Parameters}
\label{sec:ndparams}

\begin{figure}[t]
    \centering  
    \includegraphics{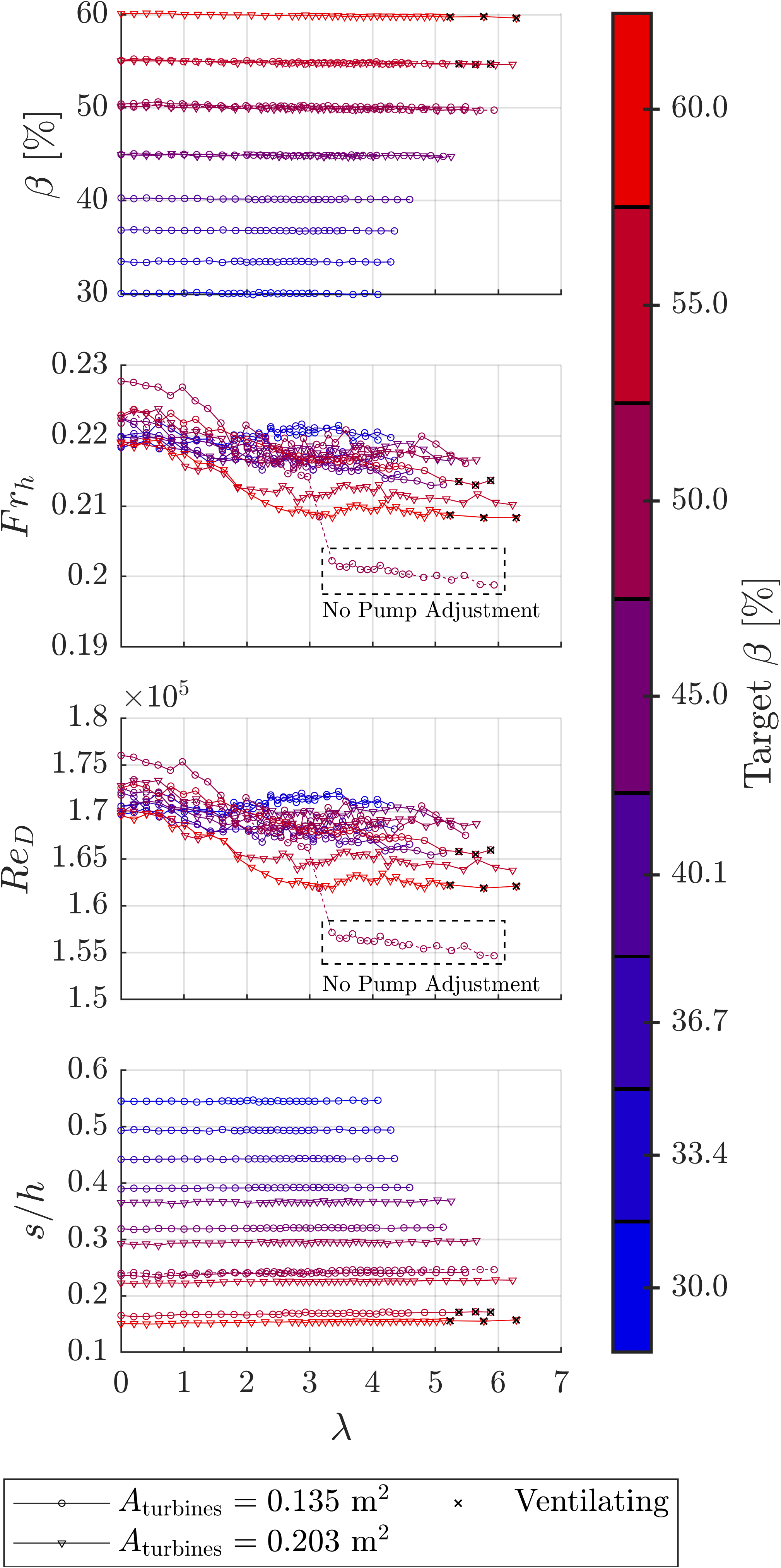}
    \caption{Time-averaged non-dimensional flow parameters as measured during the experiments. The marker of each line indicates the $A_{\mathrm{turbines}}$ used to achieve a particular $\beta$. Test points at which ventilation occurred are marked with `x'. The dashed box indicates the step-change in $Fr_h$ and $Re_D$ observed for the $\beta = 50.0\%$, $A_{\mathrm{turbines}} = 0.135 \ \mathrm{m}^2$ case without pump frequency adjustment.}
    \label{fig:ndParams}
\end{figure}

The time-averaged measured values of $\beta$, $Fr_h$, $Re_D$, and $s/h$ at each tip-speed ratio are shown in \cref{fig:ndParams}. Across all experiments, the measured $\beta$ are within $1.5\%$ of the target values in \Cref{tab:expMatrix}, and the measured $Fr_h$ and $Re_D$ do not deviate more than $5\%$ from the nominal values in \Cref{tab:expMatrix}. However, all of these parameters vary slightly with $\lambda$ due to turbine-channel interactions. As $\lambda$ increases, the array presents greater resistance to the flow, causing a reduction of the upstream freestream velocity and a rise in the upstream free surface, followed by a drop in the free surface across the turbines as the flow accelerates through the rotors. This causes $\beta$, $Fr_h$ and $Re_D$ as measured upstream of the turbine to decrease slightly with $\lambda$, and $s/h$ as measured upstream of the turbine to increase slightly with $\lambda$. Most of the variation in these parameters associated with turbine-channel interaction  occurs for $\lambda < 2$ (\cref{fig:ndParams}).
Rotor ventilation occurred only for the highest tip-speed ratios at the highest $\beta$ for each $A_{\mathrm{turbines}}$, during which the turbine blades pierced the free surface during their downstream sweep. The turbine-channel interaction did not affect the turbulence intensity, which was ${\sim}2\%$ for all test conditions.

During operation, the actual $h$ and $U_{\infty}$ in the flume are a function of the volume of water in the flume, the pump drive frequency, and the resistance to flow imposed by the turbines (primarily a function of their rotation rate). It is theoretically possible to hold $Fr_h$ and $Re_D$ truly constant across all cases by adjusting the static water depth or pump drive frequency at each $\beta\!-\!\lambda$ set-point to compensate for the array's effect on the flow. However, the trial-and-error iteration required to achieve this would be  experimentally intractable. Therefore, for each target blockage, we choose to set the flume fill and pump frequency to that which achieves the nominal $h$ and $U_{\infty}$ when no turbines are present, and report the variation in these non-dimensional parameters during experiments as in \cref{fig:ndParams}. 
However, for the case of $\beta = 50.0\%$ and $A_{\mathrm{turbines}} = 0.135 \ \mathrm{m}^2$ a step-change in $U_{\infty}$ was observed near $\lambda = 3$, resulting in corresponding sharp decreases in $Fr_h$ and $Re_D$ (indicated by the dashed black boxes in \cref{fig:ndParams}). As this step-change was repeatable and only observed for this configuration, we attribute this to a unique interaction between the flume and the turbines under these conditions. Consequently, for this case only, the pump drive frequency was increased halfway through the test to counteract the velocity reduction and maintain similar $Fr_h$ and $Re_D$ to that measured in the rest of the experiments (\cref{fig:ndParams}).

\subsection{Array Performance}

\begin{figure*}
    \centering
    \includegraphics{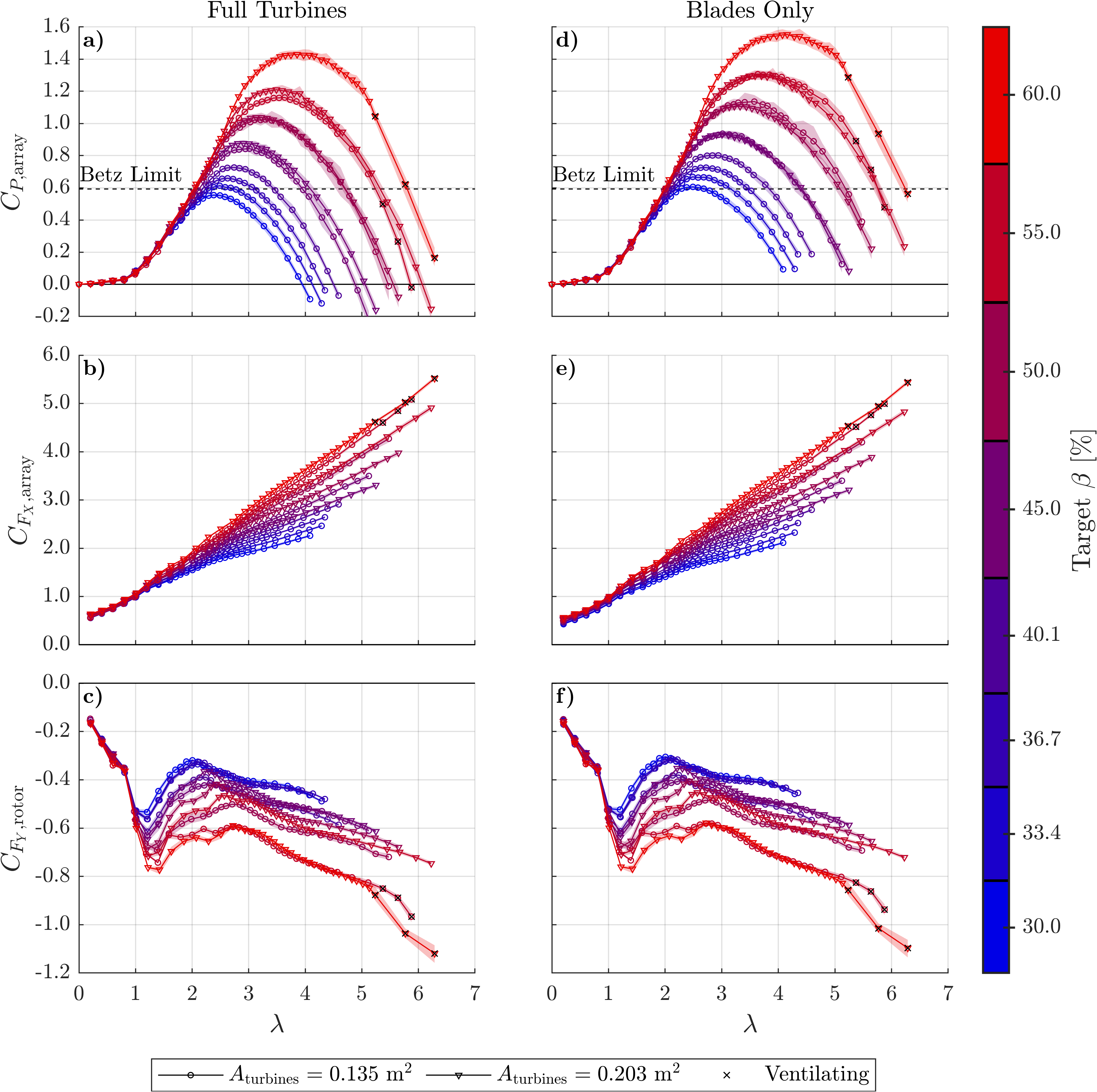}
    \caption{Time-averaged $C_{P,\mathrm{array}}$, $C_{F_{X},\mathrm{array}}$, and $C_{F_{Y},\mathrm{rotor}}$ as a function of $\lambda$ for the full-turbines (left column) and blades only (right column). The marker of each line indicates the $A_{\mathrm{turbines}}$ used to achieve a particular $\beta$. The shaded regions indicate the interquartile range of the array- and cycle-averaged performance at each $\beta$ and $\lambda$ (the vertical span of the shaded region at each point is similar to the size of the plot markers). Test points at which ventilation occurred are marked with `x'.}
    \label{fig:allPerf}
\end{figure*}

\cref{fig:allPerf}a shows the time-averaged array-average efficiencies as a function of $\beta$ and $\lambda$.
In agreement with prior work, $C_{P,\mathrm{array}}$ tends to increase as the array blockage ratio is increased. This trend is primarily observed for $\lambda > 1.5$; at lower $\lambda$, $C_{P,\mathrm{array}}$ does not vary significantly with blockage. Additionally, as $\beta$ increases the array produces power over a broader range of tip-speed ratios, and the $\lambda$ at which maximum $C_{P,\mathrm{array}}$ occurs increases.
Beginning at $\beta = 33.4\%$, $C_{P,\mathrm{array}}$ exceeds the Betz limit \citep{burton_wind_2011} and, at $\beta = 55.0\%$, values of $C_{P,\mathrm{array}}$ begin to exceed unity. Such efficiencies are not violations of energy conservation since the definition of $C_{P}$ given in \eqref{eq:cp} considers only the kinetic power that passes through the rotor plane. This definition neglects the available power associated with the fluid's potential energy, which is appreciably drawn down as $\beta$ and thrust increase. Given the relevance of both the freestream kinetic and potential energy for high-confinement arrays, a more representative efficiency metric may resemble the hydraulic efficiency of a hydropower turbine, in which the available power is a function of  volumetric flow rate and net head. However, for comparison with prior studies, the conventional definition of $C_P$ is used here. 

Time-averaged $C_{F_{X},\mathrm{array}}$ and $C_{F_{Y},\mathrm{rotor}}$ are shown in \cref{fig:allPerf}b and \cref{fig:allPerf}c, respectively. 
As expected from theory and supported by prior work, the array-average thrust coefficient increases as the blockage ratio is increased. 
Similarly, the magnitude of the lateral force coefficient (which is seldom reported for cross-flow turbines in the literature) also tends to increase with $\beta$. 
As for efficiency, $C_{F_{X},\mathrm{array}}$ and $C_{F_{Y},\mathrm{rotor}}$ do not vary significantly with blockage at low tip-speed ratios ($\lambda \leq 1$).

As mentioned in \Cref{sec:ndparams}, ventilation of the turbine rotors occurred only at high $\lambda$ for the highest $\beta$ tested for each $A_{\mathrm{turbines}}$. At these test points, which are well beyond the maximum efficiency point, foil drag due to ventilation amplifies the decreases in $C_{P,\mathrm{array}}$ and increases in $C_{F_{X},\mathrm{array}}$ and $C_{F_{Y},\mathrm{rotor}}$.

\begin{figure*}[t]
    \centering
    \includegraphics{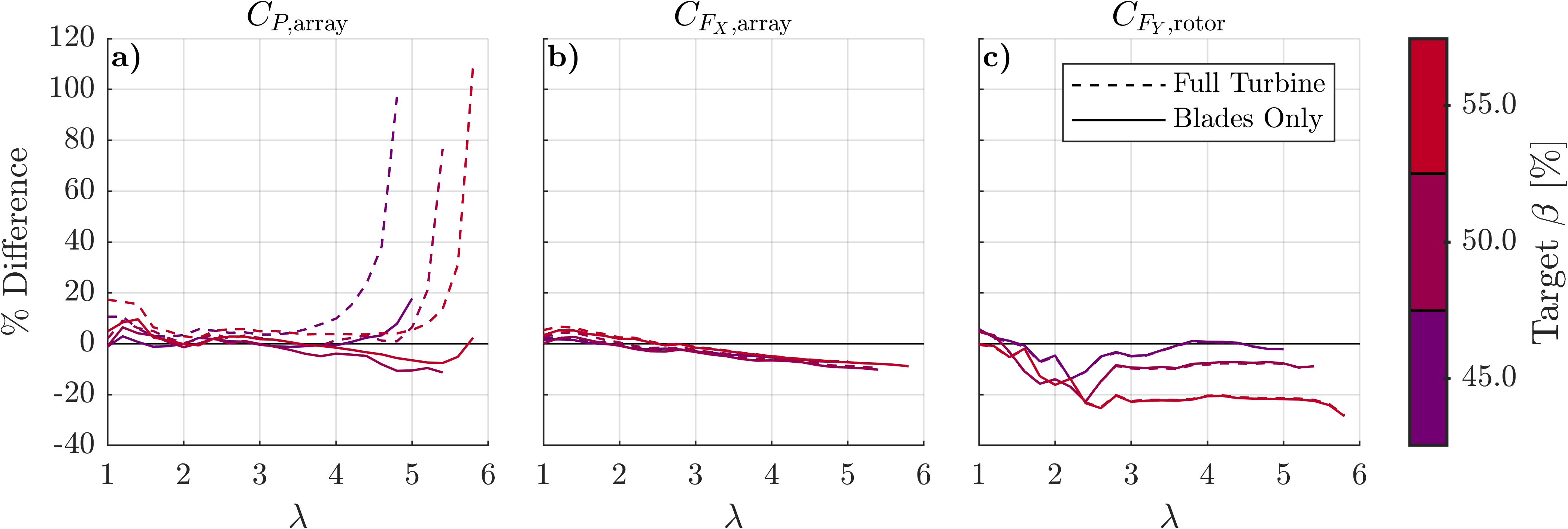}
    \caption{Percent difference in performance and force coefficients (relative to the mean values in \cref{fig:allPerf}) between the $A_{\mathrm{turbines}} = 0.135 \ \mathrm{m}^2$ case and the $A_{\mathrm{turbines}} = 0.203 \ \mathrm{m}^2$ case at 45.0\%, 50.0\%, and 55.0\% blockage. A positive percent difference implies that the $A_{\mathrm{turbines}} = 0.203 \ \mathrm{m}^2$ case performed better than the $A_{\mathrm{turbines}} = 0.135 \ \mathrm{m}^2$ case. While the percent difference in $C_{P,\mathrm{array}}$ does tend to increase with $\lambda$, the sharp increase near the end of each full turbine curve is a result of division by small $C_{P,\mathrm{array}}$ values when computing the percent difference.}
    \label{fig:turbBladeDeltas}
\end{figure*}

\subsection{Blade-level performance}
\label{sec:turbVsBlade}

For $\beta = 45.0\%$, $50.0\%$, and $ 55.0\%$, the results in \cref{fig:allPerf}a-c show that even when $Fr_h$ and $Re_D$ are held constant, achieving the same $\beta$ via different values of $A_{\mathrm{turbines}}$ and $A_{\mathrm{channel}}$ yields similar, but not identical, performance. 
We attribute this to the relative difference in support structure losses and forces between the two cases. The $A_{\mathrm{turbines}} = 0.135 \ \mathrm{m}^2$ configuration uses shorter blades than the $A_{\mathrm{turbines}} = 0.203 \ \mathrm{m}^2$ configuration, but identically sized blade support structures and fixturing. Therefore, any parasitic torque or drag forces associated with these supports are normalized by a smaller projected area when power and force coefficients are calculated. This results in lower power coefficients and higher force coefficients for the $A_{\mathrm{turbines}} = 0.135 \ \mathrm{ m}^2$ configuration than for the $A_{\mathrm{turbines}} = 0.203 \ \mathrm{ m}^2$ configuration.

To account for disparities in $C_{P,\mathrm{array}}$ between the different configurations tested at $\beta = 45.0\%$, $50.0\%$, and $ 55.0\%$, we subtract support structure losses ($C_{P,\mathrm{supports}}$) via \eqref{eq:cpBlade} to estimate the $C_{P,\mathrm{array}}$ associated with the blades only, the results of which are given in \cref{fig:allPerf}d. The agreement in full turbine and blade-only $C_{P,\mathrm{array}}$ between the two $A_{\mathrm{turbines}}$ configurations at each blockage is shown in \cref{fig:turbBladeDeltas}a. For $\beta = 45.0\%$, subtracting $C_{P,\mathrm{supports}}$ improves $C_{P,\mathrm{array}}$ agreement between the two $A_{\mathrm{turbines}}$ configurations at all $\lambda$. For $\beta = 50.0\%$, subtracting $C_{P,\mathrm{supports}}$ improves agreement in $C_{P,\mathrm{array}}$ between the two $A_{\mathrm{turbines}}$ configurations up to $\lambda\approx 3$, but agreement worsens for $\lambda > 3$. Similarly, for $\beta = 55.0\%$, subtracting $C_{P,\mathrm{supports}}$ improves agreement in $C_{P,\mathrm{array}}$ up to $\lambda \approx 3.5$, but agreement worsens for $\lambda > 3.5$. 

We attribute the poorer agreement in blade-only $C_{P,\mathrm{array}}$ at high $\beta$ and $\lambda$ to the difficulty of estimating representative $C_{P,\mathrm{supports}}$ in the absence of the turbine blades. 
As described in \Cref{sec:perfMet}, $C_{P,\mathrm{supports}}$ is estimated by testing an array of bladeless turbines at the same nominal $Fr_h$, $s/h$, and $Re_D$ as the array of full turbines. However, the array of bladeless turbines does not influence the flow field in the same way that the array of full turbines does. Specifically, as $\beta$ and $\lambda$ are increased, the thrust on the array increases, resulting in a free surface drop across the rotors and acceleration of the flow bypassing the array. These changes to the flow field are absent for an array of bladeless turbines. 
Consequently, we hypothesize that the superposition technique in \eqref{eq:cpBlade} breaks down at higher $\beta$ and $\lambda$, where turbine-channel interactions are most significant. Despite these limitations, it is remarkable that \eqref{eq:cpBlade} improves $C_{P,\mathrm{array}}$ agreement at \textit{any} $\lambda$ for $\beta = 45.0\%$, $50.0\%$, and $ 55.0\%$ given that appreciable turbine-channel interactions are observed at these blockage ratios for lower values of $\lambda$.

Blade-only $C_{F_{X},\mathrm{array}}$ and $C_{F_{Y},\mathrm{rotor}}$ are similarly estimated via analogous equations to \eqref{eq:cpBlade}, the results of which are shown in \cref{fig:allPerf}e and \cref{fig:allPerf}f. Unlike for $C_{P,\mathrm{array}}$, subtracting $C_{F_{X},\mathrm{supports}}$ and $C_{F_{Y},\mathrm{supports}}$ does not meaningfully change the agreement in $C_{F_{X},\mathrm{array}}$ or $C_{F_{Y},\mathrm{rotor}}$ between the two $A_{\mathrm{turbines}}$ configurations at $\beta = 45.0\%$, $50.0\%$, or $ 55.0\%$ for any $\lambda$, implying that the forces on the blades dominate the forces on the support structures.
Additionally, as highlighted in \cref{fig:turbBladeDeltas}b and \cref{fig:turbBladeDeltas}c, the difference in the force coefficients between the two $A_{\mathrm{turbines}}$ configurations does not change significantly at high $\lambda$ (unlike  $C_{P,\mathrm{array}}$). As before, we hypothesize that the lack of force coefficient agreement between turbines with different aspect ratios is due to differences between the flow fields experienced by the support structures when blades are present versus when blades are absent. Further investigation into techniques for estimating blade-only force coefficients is warranted, but is outside the scope of the present study.

\subsection{Evaluation of approaches for varying blockage}

As both of the blockage-varying approaches from \Cref{sec:background} were utilized in the present experiments, we now consider the advantages and disadvantages of implementing each approach.

Variation of $A_{\mathrm{channel}}$ via changing the water depth was used to achieve $\beta = 30.0\% - 55.0\%$ with $A_{\mathrm{turbines}} = 0.135 \ \mathrm{m}^2$ and $\beta = 45.0\% - 60.0\%$ with $A_{\mathrm{turbines}} = 0.203 \ \mathrm{m}^2$. The blockages testable with each $A_{\mathrm{turbines}}$ were constrained by the relative sizes of the flume and the turbines, as well as the physical limitations of the test facility. In the Tyler flume, the minimum testable blockage for each $A_{\mathrm{turbines}}$ was set by the maximum dynamic channel depth of 0.60 m, above which overtopping of the flume walls occurs. The maximum testable blockage was constrained by the ventilation risk associated with low $s/h$: the lower the water depth, the higher the $\beta$ that can be achieved, but the closer the turbines are to the surface the greater the risk of ventilating. However, as described in \Cref{sec:background}, the range of testable blockages was further constrained to the water depths at which $Fr_h$ and $Re_D$ could be matched across all tests via corresponding adjustments to the freestream velocity and water temperature. Critically, temperature control is required to avoid convolving blockage effects with variations in the Reynolds number; as many flumes do not have this capability, this is a general limitation of the variable-$A_{\mathrm{channel}}$ approach unless Reynolds-independent performance can be achieved. In the present study, the relatively wide range of temperatures achievable in the Tyler flume enabled careful control of non-dimensional flow parameters and an effective isolation of blockage effects. Even so, the duration of each experiment was extended by the need to adjust the flume fill and water temperature for each test.

Given the facility requirements of the variable-$A_{\mathrm{channel}}$ approach, the present experiments also explored how $\beta$ could be varied at fixed $A_{\mathrm{channel}}$ through variation in $A_{\mathrm{turbines}}$. For example, as shown in \Cref{tab:expMatrix}, array blockage ratios of 40.1\% and 60.0\% were achieved simply by testing arrays with $A_{\mathrm{turbines}} = 0.135 \ \mathrm{ m}^2$ and $A_{\mathrm{turbines}} = 0.203 \ \mathrm{ m}^2$ at the same nominal water depth (and thus same nominal $A_{\mathrm{channel}}$). Since the water depth was unchanged, testing different blockages was convenient and fast since no adjustments to the freestream velocity or temperature were necessary. In a general case, if no secondary effects are introduced by changing $A_{\mathrm{turbines}}$, then the range of testable blockages at a given facility using this approach would be constrained only by 1) the minimum and maximum blade spans available, and 2) ventilation risk at the highest blockages. 

However, as shown in \cref{fig:allPerf}, turbines tested at similar $\beta$, $Fr_h$, and $Re_D$, but different $A_{\mathrm{turbines}}$ and $A_{\mathrm{channel}}$, exhibit small, but appreciable differences in power and force coefficients. The superposition-based support structure subtraction techniques for reconciling these differences begin to break down for $C_{P,\mathrm{array}}$ at high $\beta$ and $\lambda$, and do not have any effect on disagreements in $C_{F_{X},\mathrm{array}}$ or $C_{F_{Y},\mathrm{rotor}}$ across $A_{\mathrm{turbines}}$. While, for these experiments, the disparities in $C_{P,\mathrm{array}}$, $C_{F_{X},\mathrm{array}}$ and $C_{F_{Y},\mathrm{rotor}}$ between the different $A_{\mathrm{turbines}}$ are small relative to the overall blockage effects, the scale of these disparities likely depends on the specific blade spans and support structures used \citep{strom_impact_2018, hunt_effect_2020, villeneuve_increasing_2021}. Consequently, interpreting trends in blockage effects obtained via a variable-$A_{\mathrm{turbines}}$ approach requires better models for support structure effects than simple experimental superposition. The uncertainty associated with the variable-$A_{\mathrm{turbines}}$ approach can only be quantified by changing $A_{\mathrm{turbines}}$ at constant $\beta$, as was performed in this study at $\beta = 45.0\%$, $50.0\%$, and $55.0\%$. To do so while holding $Fr_h$ and $Re_D$ constant requires either a facility with temperature control or a facility capable of velocities of ${\sim}10 $ m/s to achieve Reynolds independence. Consequently, the facility requirements for fully interpreting results obtained via the variable-$A_{\mathrm{turbines}}$ method negate the principal advantage of this method. Therefore, the variable-$A_{\mathrm{channel}}$ fixed-$A_{\mathrm{turbines}}$ approach is most robust.

\section{Conclusion}
In this work, we explored two experimental methods for characterizing the effects of blockage ratio on the performance of an array of two, straight-bladed cross-flow turbines operating in a water channel. For fixed $A_{\mathrm{turbines}}$ and variable $A_{\mathrm{channel}}$, the blockage ratio is most easily varied by changing the water depth, with corresponding changes in the freestream velocity and temperature to hold the Reynolds number and Froude numbers constant. For fixed $A_{\mathrm{channel}}$ and variable $A_{\mathrm{turbines}}$, the blockage ratio is most appropriately varied by changing the blade span, as several secondary effects are introduced if the turbine diameter or number of turbines are changed. A laboratory-scale array operating at blockages between 30\% and 60\% is tested using both approaches. While similar trends in efficiency and force are observed regardless of approach, the values of the array-average performance and force coefficients vary with the method used to achieve a particular $\beta$. For future experimental studies focusing on blockage effects, we recommend that the blockage ratio is varied by changing $A_{\mathrm{channel}}$ with fixed $A_{\mathrm{turbines}}$ while holding the Reynolds and Froude numbers constant. Although this method requires the use of a flume with temperature control or the ability to achieve Reynolds-invariant turbine performance, we find that it was the most robust approach since changes in blade-support structure interactions associated with changes in $A_{\mathrm{turbines}}$ can be difficult to quantify.

We recommend that future studies investigate more robust methods for estimating the parasitic losses and drag forces of turbine blade support structures such that blade-only efficiency and force coefficients can be better estimated at high blockage and tip-speed ratio, allowing more accurate comparisons of blockage effects to be drawn across turbines. The development of such a method would improve the reliability of the fixed-$A_{\mathrm{channel}}$ variable-$A_{\mathrm{turbines}}$ experimental approach, and facilitate studies of blockage effects at a wider range of test facilities. Additionally, as ventilation was a constraint on experimental design for both experimental methods, future work should evaluate how the onset and effects of ventilation are influenced by the blockage ratio, the Froude number, and the normalized submergence depth. 


%




\section*{Acknowledgement}
The authors would like to thank Gregory Talpey and Gemma Calandra for their assistance in commissioning the high-blockage test-rig, as well as help with data collection. The authors would also like to thank Abigale Snortland for several insightful discussions regarding support structure torque and force subtraction techniques.

\ifCLASSOPTIONcaptionsoff
  \newpage
\fi




\balance

\AtNextBibliography{\footnotesize}    
\printbibliography


\end{document}